\documentclass[a4paper,fleqn,usenatbib]{mnras}
\usepackage{newtxtext,newtxmath}

\usepackage[T1]{fontenc}
\usepackage{ae,aecompl}

\usepackage{amsmath}	
\usepackage{amssymb}	
\usepackage{graphicx}	

\usepackage{natbib}
\usepackage{longtable}
\usepackage{color}
\newcommand{\asec}{$^{\prime\prime}$}

\def\r1415{$^{14}$N/$^{15}$N}
\def\HOCO{HOCO$^{+}$}

\def\H{N$_{2}$H$^{+}$}

\def\15N{$^{15}$NNH$^+$}
\def\N15{N$^{15}$NH$^+$}

\def\HCOp{\mbox{HCO$^+$}}
\def\HCOpI{\mbox{H$^{13}$CO$^+$}}

\def\METH{CH$_3$OH}
\def\HII{H{\sc ii}}

\def\kms{\mbox{km~s$^{-1}$}}

\def\cmq{cm$^{-2}$}

\def\Tex{\mbox{$T_{\rm ex}$}}

\def\TMB{\mbox{$T_{\rm MB}$}}

\title[\HOCO\ in massive star-forming clumps]{Protonated CO$_2$ in massive star-forming clumps}

\author[Fontani et al.]{F. Fontani,$^{1}$\thanks{E-mail: fontani@arcetri.astro.it}
            A. Vagnoli,$^{2}$
            M. Padovani,$^{1}$
            L. Colzi,$^{1,2}$
            P. Caselli,$^{3}$
            and V.M. Rivilla$^{1}$
          \\
$^{1}$INAF-Osservatorio Astrofisico di Arcetri, Largo E. Fermi 5, I-50125, Florence, Italy \\
$^{2}$Dipartimento di Fisica e Astronomia, Universit\`a degli Studi di Firenze, I-50125 Firenze, Italy \\
$^{3}$Centre for Astrochemical Studies, Max-Planck-Institute for Extraterrestrial Physics, Giessenbachstrasse 1, 85748 Garching, Germany  \\
          }

\date{Accepted XXX. Received YYY; in original form ZZZ}

\pubyear{2017}

\begin{document}
\label{firstpage}
\pagerange{\pageref{firstpage}--\pageref{lastpage}}
\maketitle

\begin{abstract}
Interstellar CO$_2$ is an important reservoir of carbon and oxygen,
and one of the major constituents of the icy mantles of dust grains, but it is not observable 
directly in the cold gas because has no permanent dipole moment. Its protonated form, \HOCO , is believed 
to be a good proxy for gaseous CO$_2$. However, it has been detected in only a few star-forming regions 
so far, so that its interstellar chemistry is not well understood. We present new 
detections of \HOCO\ lines in 11 high-mass star-forming clumps. Our observations 
increase by more than three times the number of detections in star-forming regions so far. 
We have derived beam-averaged 
abundances relative to H$_2$ in between 
$0.3$ and $3.8\times 10^{-11}$. We have compared these values with 
the abundances of \HCOpI, a possible gas-phase precursor of \HOCO, and 
\METH, a product of surface chemistry. We have found a positive correlation with 
\HCOpI, while with \METH\ there is no correlation. 
We suggest that the gas-phase formation route starting from \HCOp\ plays an important role in 
the formation of \HOCO, perhaps more relevant than protonation of CO$_2$ 
(upon evaporation of this latter from icy dust mantles). 
\end{abstract}


%
\section{Introduction}
\label{intro}

Carbon dioxide (CO$_2$) is a relevant molecular species in a variety of interstellar
environments. In comets, planetary atmospheres, and interstellar ices, its abundance
is a significant fraction ($\sim 0.1 - 0.5$) of that of water (e.g.~Bergin et al.~\citeyear{bergin}, 
Whittet et al.~\citeyear{whittet}, McKay et al.~\citeyear{mckay}, Hoang et al.~\citeyear{hoang}).
CO$_2$ ice is one of the main constituent of the icy mantles of
dust grains (\"Oberg et al.~\citeyear{oberg}). In the gas-phase, CO$_2$ can be observed 
directly through ro-vibrational transitions (e.g.~van Dishoeck et al.~\citeyear{vandishoeck}), 
but the lack of permanent dipole moment, and hence of a pure rotational spectrum, makes it 
impossible a detection in cold environments. Instead, 
its protonated form, \HOCO, has been detected towards the Galactic center (Thaddeus 
et al.~\citeyear{thaddeus}, Minh et al.~\citeyear{minh}, Neill et al.~\citeyear{neill}), in
diffuse and translucent clouds (Turner et al.~\citeyear{turner}), but only in a handful 
of star-forming regions: in the low-mass pre-stellar core L1544 (Vastel et al.~\citeyear{vastel}), 
in the protostars L1527 and IRAS 16293--2422 (Sakai et al.~\citeyear{sakai}, Majumdar et
al.~\citeyear{majumdar}), and in the protostellar shock L1157--B1 (Podio et al.~\citeyear{podio}). 

In cold and dense gas, two main chemical formation pathways have been proposed: {\bf (1)} a gas-phase 
route from the reaction ${\rm HCO^+ + OH \leftrightarrow HOCO^+ + H }$, and {\bf (2)} the 
protonation of CO$_2$ (mainly upon reaction with H$_3^+$) desorbed from grain mantles 
(see e.g.~Vastel et al.~\citeyear{vastel}, Bizzocchi et al.~\citeyear{bizzocchi}). 
In scenario {\bf (1)}, CO$_2$ would be a product of \HOCO\ (after dissociative recombination), 
while the opposite is expected in scenario {\bf (2)}.  Due to the lack of stringent 
observational constraints, it is unclear yet which of these two mechanisms is dominant, and
under which physical conditions. 
Constraining the abundance of \HOCO\ has important implications also for the abundance 
of CO$_2$ in ice. In fact, if \HOCO\ is formed in the cold gas and then, upon dissociative
recombination, gives rise to CO$_2$, this latter could freeze-out on grain mantles and contribute 
to the amount of CO$_2$ ice observed in dark clouds
(Bergin et al.~\citeyear{bergin}), although this cannot explain the large amount of solid CO$_2$
measured along the line of sight of background stars (Boogert et al.~\citeyear{boogert})
or deeply embedded massive young stars (van Dishoeck et al.~\citeyear{vandishoeck}). 
In fact, the formation of CO$_2$ ice from surface reactions is still debated. Laboratory 
experiments suggested formation of CO$_2$ ice from 
${\rm CO + O \rightarrow CO_2}$ (D'Hendecourt et al.~\citeyear{dhendecourt}), which 
however needs a strong UV irradiation, and hence it is expected to be inefficient in dark clouds. 
Other surface reactions have been proposed, such as cosmic-ray bombardment on carbonaceous 
grains covered by water ice (Mennella et al.~\citeyear{mennella}), or the radical-radical reaction 
${\rm OH + CO \rightarrow CO_2 + H}$ (Garrod \& Pauly~\citeyear{gep}, Ioppolo et al. 2011, 
Noble et al. 2011). However, such process involves the diffusion of heavy radicals, difficult to happen at 
dust temperatures below $\sim 30$~K, although high precision measurements are not available yet, 
and new promising techniques have been proposed (e.g.~Cooke et al.~\citeyear{cooke}) 
to shed light on this important surface chemistry process.

In this paper, we present new detections of \HOCO\ in 11 high-mass star-forming regions, 
belonging to an evolutionary sample of 27 clumps divided into the three main evolutionary 
categories of the massive star formation process (Fontani et al.~\citeyear{fontani2011}): high-mass
starless cores (HMSCs), high-mass protostellar objects (HMPOs) and Ultra-compact \HII\
regions (UCHIIs). The sample has been extensively observed in several dense
gas tracers, with the aim of studying the chemical evolution of these molecules during
the massive star-formation process
(Fontani et al.~\citeyear{fontani2011},~\citeyear{fontani2014},~\citeyear{fontani2015a},~\citeyear{fontani2015b},
~\citeyear{fontani2016},
Colzi et al.~\citeyear{colzi},~\citeyear{colzib}, Mininni et al.~\citeyear{mininni}).
This work represents the first
study of protonated carbon dioxide in a statistically relevant number of star forming regions.

\section{Observations}
\label{obs}

The spectra analysed in this work are part of the dataset presented in Colzi et al.~(\citeyear{colzi}). 
These data were obtained with the IRAM-30m Telescope in the 3~mm band with the EMIR receiver,
covering the frequency ranges 85.31 -- 87.13~GHz and 88.59 -- 90.41~GHz, toward 26 dense cores in 
massive star-forming regions divided in the 3 evolutionary categories HMSCs, HMPOs and UCHIIs
(see Sect.~\ref{intro}).
For details on the source selection, see Fontani et al.~(\citeyear{fontani2011}).
The atmospheric conditions were very stable, with amounts of precipitable water vapour
in the range 3 -- 8~mm. We observed in wobbler-switching mode (wobbler throw of 240\asec ). 
Pointing was checked almost every hour on nearby quasars, planets, or bright HII regions.
The data were calibrated with the chopper wheel technique (see Kutner \& Ulich~\citeyear{keu}), 
and the calibration uncertainty is estimated to be about $10\%$. More details are given
in Colzi et al.~(\citeyear{colzi}).
The spectra have been reduced and analysed with the software CLASS of the GILDAS package.
The detected lines have been fit with a gaussian shape.

\section{Results}
\label{res}

\begin{figure}
\begin{center}
\includegraphics[width=8.5cm,angle=0]{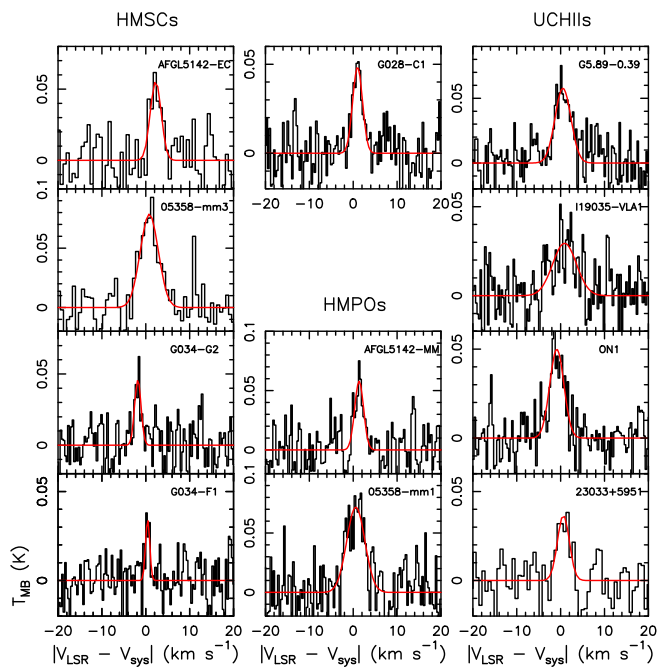}
 \caption{Spectra of the \HOCO\ ($4_{0,4}-3_{0,3}$) line in the 11 detected sources.
 The y-axis is in \TMB\ units. The x-axis corresponds to $|V_{\rm LSR}-V_{\rm sys}|$,
 i.e. the difference between the local standard of rest velocity, $V_{\rm LSR}$, and the nominal
 systemic velocity, $V_{\rm sys}$, given in Fontani et al.~\citeyear{fontani2011}.
 The red curves superimposed on the spectra represent the best gaussian fits.}
 \label{spectra}
\end{center}
\end{figure}

\begin{figure}
\begin{center}
{\includegraphics[width=7cm,angle=0]{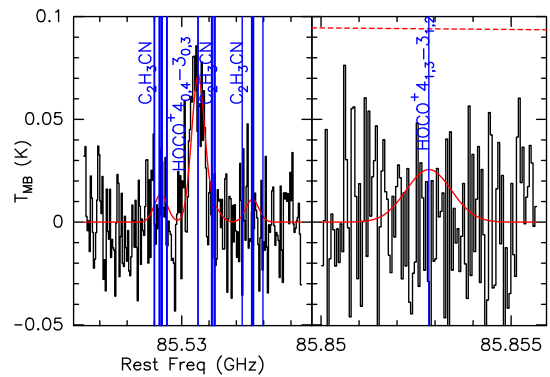}}
 \caption{{\it Left panel}: spectrum of the \HOCO\ $4_{0,4}-3_{0,3}$ line towards 05358--mm1
 superimposed on the synthetic spectrum (in red) calculated with MADCUBA.
 The synthetic spectrum is calculated for a \HOCO\ column density of 
$4.3 \times 10^{12}$ \cmq\ and a C$_2$H$_3$CN column density of $10^{18}$ \cmq,
 which is the value measured towards Sgr B2(N) (see Sect.~\ref{res}). The rest
 frequencies of the nearby C$_2$H$_3$CN transitions are indicated by vertical lines.
 {\it Right panel}: spectrum of 05358--mm1 centred on the rest frequency of 
 \HOCO\ $4_{1,3}-3_{1,2}$, superimposed on the synthetic spectrum created with 
 the same parameters used in Left panel: the expected peak of the line is well 
 under the 3$\sigma$ rms noise (shown by the red dashed horizontal line).}
 \label{synthetic}
\end{center}
\end{figure}

Table~1 lists the 27 observed sources: 11 HMSCs, 9 HMPOs, and 7 UCHIIs.
We have detected clearly (signal-to-noise ratio $\geq 5$) the \HOCO\ $4_{0,4}-3_{0,3}$ transition 
(\HOCO\ 4--3 hereafter, $E_{\rm up}\sim 10.3$~K) at 85531.497~MHz (Bizzocchi et 
al.~\citeyear{bizzocchi}) in five HMSCs, two HMPOs and four UCHIIs. The spectra of the 
detected sources are shown in Fig.~\ref{spectra}. The relatively high detection ($\sim 50\%$) 
in the HMSCs indicates that \HOCO\ is a species abundant in cold and dense gas.
The detection rate decreases during the HMPO phase ($\sim 22\%$), and then it increases
again at the UCHII stage ($\sim 57\%$). 
Since we have detected only one transition of \HOCO, to confirm that the detected line
is indeed \HOCO\ and rule out contamination by transitions of other molecules, we have 
simulated the spectrum of one of the two sources in which the line has the highest intensity
peak, namely 05358-mm1 (Fig.~\ref{spectra}), 
and used the software MADCUBA\footnote{Madrid Data Cube Analysis on image is a software 
to visualize and analyse astronomical single spectra and datacubes (Mart\'in et al., in prep.,
Rivilla et al.~\citeyear{rivilla}).} to search for emission of nearby lines potentially blended with \HOCO\ 4--3. 
Belloche et al.~(\citeyear{belloche}) have shown that in Sgr B2(N) the only species 
with lines that could contaminate \HOCO\ 4--3 is C$_2$H$_3$CN. 
They found C$_2$H$_3$CN column densities above $10^{18}$ \cmq\ and 
line widths larger than those that we have found for \HOCO. 
To simulate the spectrum, we have assumed as excitation temperature the kinetic 
temperature obtained from ammonia (39 K, Fontani et al.~\citeyear{fontani2011}) and 
the same line width obtained fitting \HOCO\ 4--3 (4.9 \kms). As shown in Fig.~\ref{synthetic}
(top panel), even assuming a huge column density of $10^{18}$ \cmq\ as in Sgr B2(N), the 
expected C$_2$H$_3$CN line intensities are well below the spectral rms, and well
separate from the \HOCO\ 4--3 line. Another transition of \HOCO, namely the $4_{1,3}-3_{1,2}$ line 
centred at 85852.8576 MHz ($E_{\rm u}\sim 48$~K), falls in our band but it is undetected. 
Because its line strength and Einstein coefficient are quite similar to those of the 
$4_{0,4}-3_{0,3}$ transition (see Bizzocchi et al.~\citeyear{bizzocchi}),
we have investigated if its non-detection is due to the lack of sensitivity. Fig.~\ref{synthetic} 
(bottom panel) shows that indeed the synthetic spectrum of 05358--mm1 around this 
line (simulated assuming the parameters described above for \HOCO) is consistent with a 
non-detection. Because this source shows the highest intensity peak of \HOCO\ 4--3, we 
can reasonably conclude that the non-detection of the second transition is due to a sensitivity 
limit in all sources.

All detected lines are well fit by a single gaussian (Fig.~\ref{spectra}). The results
of the fit performed as explained in Sect.~\ref{obs}, namely line integrated intensity 
($\int T_{\rm MB}{\rm d}v$), peak velocity in the Local Standard of Rest (LSR)
($v_{\rm p}$), and line width at half maximum ($\Delta v$) are shown in Table~1. 
The uncertainties on $\int T_{\rm MB}{\rm d}v$ are calculated from the expression 
${\sigma}\Delta v_{\rm res}\sqrt{n}$, obtained from the propagation of errors, where
${\sigma}$ is the 1$\sigma$ root mean square noise in the spectrum, $\Delta v_{\rm res}$ the spectral
resolution, and $n$ the number of channels with signal. The uncertainties on $v_{\rm p}$ and 
$\Delta v$ are computed by the fit procedure.
The $\Delta v$ measured towards the detected UCHII regions are always larger than 
$\sim 3.2$~\kms, while they are narrower than 3~\kms\ in all HMSCs (but in 05358--mm3),
in agreement with the fact that the envelopes of UCHII regions are more turbulent than
those of HMSCs. The case of 05358--mm3 is peculiar because this core is likely externally 
heated (Fontani et al.~\citeyear{fontani2011}), hence its chemical composition and 
the emission that we observe is likely a mix between the cold and dense core nucleus 
and the warmer envelope. The line shapes are symmetric and do not show non-gaussian wings. 
This suggests that in all sources the \HOCO\ emission is associated with the bulk gas. This
finding is supported by the fact that the peak velocities are consistent with the systemic velocities 
(given in Fontani et al.~\citeyear{fontani2011}) within the uncertainties (see Fig.~\ref{spectra}). 

\subsection{\HOCO\ column densities and fractional abundances}
\label{abundances}

From the line integrated intensity, we have calculated the total column 
density of \HOCO\ assuming Local Thermodynamic Equilibrium (LTE) and optically thin conditions
from the equation:
\begin{equation}
N=\frac{8\pi \nu_{\rm ij}^3}{c^3 A_{\rm ij}}\frac{\int T_{\rm MB}{\rm d}v}{g_{\rm i}[J_{\nu}(T_{\rm ex})-J_{\nu}(T_{\rm BG})]}\frac{Q(T_{\rm ex}) {\rm exp}(E_{\rm j}/k T_{\rm ex})}{1-{\rm exp(-h\nu_{\rm ij}/k T_{\rm ex}})}\;,
\label{eq_coldens}
\end{equation}
where $\nu_{\rm ij}$ is the frequency of the transition, $A_{\rm ij}$ is the Einstein
coefficient of spontaneous emission, $g_{\rm i}$ is the statistical weight of the upper 
level, $E_{\rm j}$ is the energy of the lower level, $c$ is the speed of light, $h$ and $k$ are 
the Planck and Boltzmann constants, respectively, $T_{\rm ex}$ is the excitation temperature, $Q(T_{\rm ex})$
is the partition function computed at temperature $T_{\rm ex}$, $T_{\rm BG}$ is the background
temperature (assumed to be that of the cosmic microwave background, 2.7~K), and $J_{\nu}(T)$
is the equivalent Rayleigh-Jeans temperature (see also Eq. (A4) in Caselli et al.~\citeyear{caselli2002}).
The line strength, energy of the upper level,
and Einstein coefficient of spontaneous emission, are taken from the Cologne Database
for Molecular Spectroscopy (CDMS\footnote{https://www.astro.uni-koeln.de/cdms},
see also Bogey et al.~\citeyear{bogey}, and Bizzocchi et al.~\citeyear{bizzocchi}),
and are: $S\mu^2\sim 29.2$ D$^2$, $E_{\rm u}\sim 10.3$~K, and 
$A_{\rm ij}\sim 2.36\times 10^{-5}$ s$^{-1}$, respectively.
The assumption of optically thin emission is consistent with the low abundance of the molecule and 
with line shapes without hints of high optical depths (like, e.g., asymmetric or flat 
topped profiles). The beam averaged column densities are in the range 
$\sim 3.5\times 10^{11} - 4.6\times 10^{12}$ \cmq .
For undetected lines, we have computed the upper limits on $\int T_{\rm MB}{\rm d}v$
assuming a gaussian line with intensity peak equal to the 3$\sigma$ rms in the spectrum, 
and $\Delta v$ equal to the average value of each evolutionary group, and from this
we have derived the upper limits on $N$(HOCO$^+$) from Eq.~\ref{eq_coldens}.

We have computed the \HOCO\ fractional abundances, X[HOCO$^+$], by dividing the \HOCO\ total 
column densities by those of H$_2$, $N$(H$_2$), derived from the sub-millimeter continuum 
emission. This latter was computed from the (sub-)millimeter dust thermal continuum emission
extracted from the images of the $850$ $\mu$m survey of Di Francesco et al.~(\citeyear{difrancesco})
obtained with SCUBA at the James Clerk Maxwell Telescope (JCMT).  We have
used Eq. (A1) in Mininni et al.~(\citeyear{mininni}) to compute $N$(H$_2$), which assumes
optically thin emission, and a gas-to-dust ratio of 100. The sub-millimeter continuum fluxes,
$F_{\rm submm}$, used to compute $N$(H$_2$) have been extracted from a circular area 
equivalent to the IRAM-30m Half Power Beam Width (HPBW) at the frequency of the \HOCO\ line, 
i.e. $\sim 28$\asec. 
The uncertainty on $F_{\rm submm}$ is calculated from the propagation of errors.
The sub-mm fluxes, and the derived $N$(H$_2$) and X[HOCO$^+$] obtained as explained above, 
are given in Table~1.
For the sources not present in the survey of Di Francesco et al.~(\citeyear{difrancesco}), 
we have estimated $N$(H$_2$) following the same analysis from the APEX ATLASGAL 
continuum images (http://www3.mpifr-bonn.mpg.de/div/atlasgal/index.html) 
at $\sim 870$~GHz. We derive X[HOCO$^+$] in the range 0.3 -- 3.8$\times 10^{-11}$.
These values are intermediate between those obtained towards the pre-stellar core 
L1544 ($\sim 5\times 10^{-11}$, Vastel et al.~\citeyear{vastel}) and the hot corino IRAS 
16293--2422 ($\sim 1\times 10^{-13}$, Majumdar et al.~\citeyear{majumdar}).

\begin{table*}
\begin{center}
\caption{Results derived from gaussian fits to the lines. In cols. 2--6 we give: integrated intensity
($\int T_{\rm MB}{\rm d}v$), velocity in the Local Standard of Rest (LSR) at line peak ($v_{\rm p}$), 
full width at half maximum ($\Delta v$), assumed gas excitation temperature ($T_{\rm ex}$), and \HOCO\ beam-averaged 
total column density, $N$(HOCO$^+$), calculated as explained in Sect.~\ref{abundances}. 
The uncertainities on $N$ are obtained propagating the error
on $\int T_{\rm MB}{\rm d}v$, to which we sum a 10$\%$ of calibration error on the $T_{\rm MB}$ scale 
(see Sect.~\ref{obs}) not included in the error on $\int T_{\rm MB}{\rm d}v$. \Tex\ is assumed without 
uncertainty. Col.~7 and 8 give the sub-millimeter flux densities, $F_{\rm submm}$, and the H$_2$ 
column densities, $N$(H$_2$), derived from it, respectively (see Sect.~\ref{abundances} for details). 
The error on $N$(H$_2$) is obtained propagating the error on $F_{\rm submm}$,
to which we add a calibration error of 20$\%$ on the SCUBA absolute flux scale at 850$\mu$m
(Di Francesco et al.~\citeyear{difrancesco}). 
The \HOCO\ fractional abundances, X[\HOCO], are shown in Col.~9. Finally, in Cols. 10, 11 and 12,
we list the fractional abundances of \METH, \HCOpI, and \H\ averaged over 28\asec.}
\small
\tabcolsep 0.1cm
\begin{tabular}{cccccccccccc}
\hline \hline
Source & $\int T_{\rm MB}{\rm d}v$ & $v_{\rm p}$ & $\Delta v$ & \Tex\ & $N$(HOCO$^+$) & $F_{\rm submm}$ & $N$(H$_2$)      & X[HOCO$^+$] & X[CH$_3$OH]  & X[H$^{13}$CO$^+$] & X[N$_2$H$^+$] \\ 
           & K km s$^{-1}$ & km s$^{-1}$ & km s$^{-1}$ & K & $\times 10^{11}$ cm$^{-2}$ & Jy & $\times 10^{22}$ cm$^{-2}$            & $\times 10^{-11}$ & $\times 10^{-10}$ & $\times 10^{-11}$ &  $\times 10^{-9}$ \\
\hline 
\multicolumn{12}{c}{HMSCs} \\ 
00117--MM2  &   $\leq 0.069$  &  --  & -- & 14 & $\leq 3$ & --$^{(u)}$ & -- & -- & --  & -- & -- \\
AFGL5142--EC$^{(w)}$ & $0.17 \pm 0.03$ & $-1.6\pm 0.3$ & $2.9\pm 0.6$ & 25 & $9 \pm 2$ & 5.1$\pm 0.5$$^{(s)}$ & 11.3$\pm 3.4$      & 0.8$\pm 0.4$ & 28$\pm 8$  &  7.4$\pm 3.1$ & 2.5$\pm 1.4$ \\
05358--mm3$^{(w)}$ & $0.41 \pm 0.05$ & $-16.8\pm 0.3$ & $4.8\pm 0.6$ & 30 & $23 \pm 5$ & 6.1$\pm 0.2$$^{(s)}$ & 10.6$\pm 2.4$      & 2.2$\pm 1.0$ & 12$\pm 3$ & -- & 0.6$\pm 0.2$ \\
G034--G2 & $0.08 \pm 0.02$ & $41.8\pm 0.2$ & $1.7\pm 0.5$               & 20 & $4 \pm 1$ & 0.88$\pm 0.04$$^{(a)}$ & 4.1$\pm 1.0$         & 0.9$\pm 0.5$ & 2.2$\pm 0.5$ &  2.0$\pm 0.8$ & 1.5$\pm 0.4$\\
G034--F2 &  $\leq 0.081$  & -- & --                                                            & 20 & $\leq 4$ & 0.36$\pm 0.02$$^{(s)}$ & 1.6$\pm 0.4$             & $\leq 2.5$ & 3.1$\pm 0.8$  & 1.1$\pm 0.4$ & 1.0$\pm 0.3$ \\
G034--F1 & $0.04 \pm 0.01$ & $58.1\pm 0.2$ & $1.2\pm 0.4$                & 20 & $1.9 \pm 0.4$ & 0.27$\pm 0.02$$^{(s)}$ & 1.2$\pm 0.3$    & 1.6$\pm 1.0$ & 9.8$\pm 2.5$ & 6.7$\pm 2.7$ & 3.4$\pm 1.1$ \\
G028--C1 & $0.14 \pm 0.02$ & $79.3\pm 0.2$ & $2.6\pm 0.4$                & 17 & $6 \pm 1$ & 0.93$\pm 0.03$$^{(s)}$ & 3.6$\pm 0.8$          & 1.7$\pm 0.8$ & 3.8$\pm 0.9$ & 6$\pm 2$ & 2.3$\pm 0.6$ \\
G028--C3 &  $\leq 0.084$  & -- & --                                                           & 17 & $\leq 4$ & 0.56$\pm 0.03$$^{(a)}$ & 2.4$\pm 0.6$          & $\leq 3.5$ & -- & -- & -- \\
I20293--WC & $\leq 0.074$ & -- & --                                                         & 17 & $\leq 3$ & 1.9$\pm 0.1$$^{(s)}$ & 7.4$\pm 1.9$              & $\leq 0.4$ & 2.4$\pm 0.6$ &  4.9$\pm 1.9$ & 1.5$\pm 0.5$ \\
22134--G$^{(w)}$ &  $\leq 0.076$ & -- & --                                               & 25 & $\leq 4$ & 1.6$\pm 0.1$$^{(s)}$ & 3.6$\pm 0.9$              & $\leq 1.1$ & 4$\pm 1$ &  8.7$\pm 3.3$ & 0.4$\pm 0.1$ \\
22134--B & $\leq 0.071$ & -- & --                                                             & 17 & $\leq 3$ & 0.52$\pm 0.05$$^{(s)}$ & 2.0$\pm 0.6$           & $\leq 1.6$ & 0.9$\pm 0.3$ &  4.8$\pm 2.0$ & 0.9$\pm 0.4$ \\
\hline
\multicolumn{12}{c}{HMPOs} \\ 
00117--MM1 & $\leq 0.11$ & -- & --  &  20  & $\leq 5$ & --$^{(u)}$ & -- & -- & -- & -- & -- \\
AFGL5142--MM & $0.15 \pm 0.03$ & $-2.5\pm 0.2$ & $2.4\pm 0.4$    & 34 & $9 \pm 3$ & 6.8$\pm 0.3$$^{(s)}$ & 10.1$\pm 2.4$             & 0.9$\pm 0.5$ & 53$\pm 13$ &  10$\pm 4$ & 2.4$\pm 0.7$ \\
05358--mm1 & $0.37 \pm 0.04$ & $-17.0\pm 0.2$ & $4.9\pm 0.5$       & 39 & $25 \pm 5$ & 6.1$\pm 0.3$$^{(s)}$ & 8.4$\pm 2.1$             & 3.0$\pm 1.4$ & 30$\pm 8$ & 1.0$\pm 0.4$ & 1.4$\pm 0.04$ \\
18089--1732 & $\leq 0.13$ & -- & --                                                        &  38 & $\leq 8$ & 7.5$\pm 0.4$$^{(s)}$ & 9.6$\pm 2.4$                 & $\leq 0.8$ & 67$\pm 17$ &  14$\pm 5$ & 2.2$\pm 0.8$ \\
18517+0437 & $\leq 0.12$ & -- & --                                                        &  40 & $\leq 8$ & 6.7$\pm 0.4$$^{(a)}$ & 7.9$\pm 1.3$                & $\leq 1.0$ & 54$\pm 9$ & 21$\pm 6$ & 1.1$\pm 0.2$\\
G75--core    & $\leq 0.10$ & -- & --                                                         & 96 & $\leq 14$ & 10.0$\pm 0.4$$^{(s)}$ & 4.4$\pm 1.1$           & $\leq 3.2$ & 69$\pm 17$ & 38$\pm 14$ & 1.0$\pm 0.3$ \\
I20293--MM1  & $\leq 0.10$ & -- & --                                                     & 37 & $\leq 7$ & 3.7$\pm 0.2$$^{(s)}$ & 4.9$\pm 1.2$                 & $\leq 1.3$ & 11$\pm 3$ & 13$\pm 5$ & 9$\pm 2$ \\
I21307     & $\leq 0.087$ & -- & --                                                          & 21 & $\leq 4$ & 1.03$\pm 0.04$$^{(s)}$ & 2.9$\pm 0.7$             & $\leq 1.4$ & 4.5$\pm 1.1$ & 3.2$\pm 1.1$ & 0.8$\pm 0.3$ \\
I23385     & $\leq 0.096$ & -- & --                                                          & 37 & $\leq 6$ & 1.81$\pm 0.05$$^{(s)}$ & 2.4$\pm 0.6$             & $\leq 2.6$ & 15$\pm 4$ & 9.3$\pm 3.3$ & 0.6$\pm 0.2$ \\
\hline
\multicolumn{12}{c}{UCHIIs} \\ 
G5.89--0.39 & $0.26 \pm 0.03$ & $9.5\pm 0.2$ & $4.3\pm 0.6$      & 31 & $15 \pm 3$ & 18$\pm 1$$^{(s)}$ & 55$\pm 14$                     & 0.3$\pm 0.1$ & 8.5$\pm 2$ & 4.2$\pm 1.7$ & 0.5$\pm 0.1$ \\
I19035--VLA1 & $0.20 \pm 0.03$ & $33.3\pm 0.4$ & $6.4\pm 0.9$ & 39 & $14 \pm 3$ & 2.9$\pm 0.1$$^{(s)}$ & 3.6$\pm 0.9$               & 3.8$\pm 1.9$ & 17$\pm 4$ & 15$\pm 5$ & 5.0$\pm 1.5$ \\
19410+2336 & $\leq 0.15$  & -- & --                                                 & 21 & $\leq 7$ & 4.8$\pm 0.1$$^{(a)}$ & 13.6$\pm 3.0$                  & $\leq 0.5$ & 5.4$\pm 1.2$ & 3.6$\pm 1.2$ & 3.2$\pm 1.0$ \\
ON1  & $0.20 \pm 0.02$ & $11.2\pm 0.2$ & $3.8\pm 0.5$               & 26 & $10 \pm 2$ & --$^{(u)}$ & -- & -- & -- & -- & --  \\
I22134--VLA1 & $\leq 0.11$ & -- & --                                                & 47 & $\leq 9$ & 2.1$\pm 0.1$$^{(s)}$ & 2.1$\pm 0.5$                    & $\leq 4.2$ & 2.9$\pm 0.7$ & 2.0$\pm 0.7$ & 0.9$\pm 0.3$ \\
23033+5951 & $0.12 \pm 0.02$ & $-52.4\pm 0.3$ & $3.2\pm 0.5$ & 25 & $7 \pm 2$ & 3.5$\pm 0.1$$^{(s)}$ & 7.8$\pm 1.8$                 & 0.8$\pm 0.4$ & 5.7$\pm 1.3$ & 9.2$\pm 3.1$ & 2.6$\pm 1.4$ \\
NGC7538--IRS9 & $\leq 0.16$ & -- & --                                           & 31 & $\leq 10$ & 3.9$\pm 0.1$$^{(s)}$ & 11.9$\pm 2.7$                   & $\leq 0.8$ & 5.4$\pm 1.2$ & 4.1$\pm 1.4$ & 0.5$\pm 0.1$ \\
\hline\hline
\end{tabular}
\normalsize
\end{center}
$^{(w)}$ "warm" core having a kinetic temperature higher than 20~K, and likely externally heated (Fontani et al.~\citeyear{fontani2011});
$^{(s)}$ measured from the maps of the SCUBA survey (Di Francesco et al.~\citeyear{difrancesco});
$^{(a)}$ measured from the maps of the APEX ATLASGAL survey (http://www3.mpifr-bonn.mpg.de/div/atlasgal/index.html);
$^{(u)}$ continuum map not available either in the survey of Di Francesco et al.~(\citeyear{difrancesco}) or in the ATLASGAL survey.
\label{tab_res}
\end{table*}

\section{Discussion and conclusions}
\label{discu}

\begin{figure}
\begin{center}
\includegraphics[width=6.3cm,angle=0]{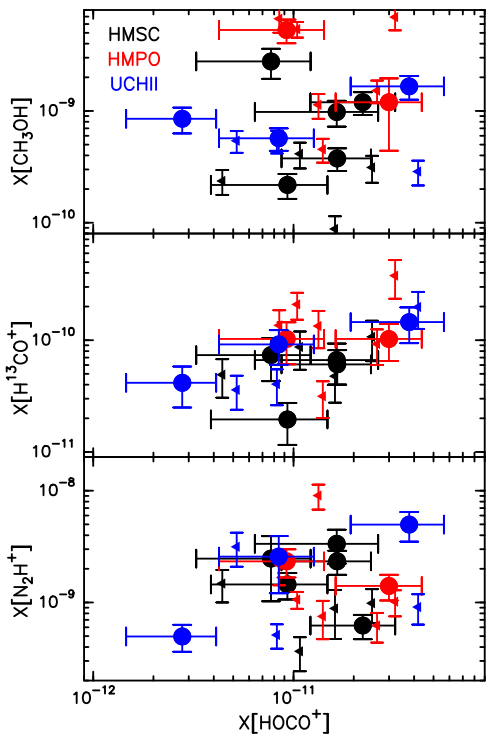}
 \caption{Abundance of \HOCO\ against that of, from top to bottom: \METH, H$^{13}$CO$^+$,
 and \H. 
 The colours indicate the different evolutionary groups as labelled in the top-left corner. 
 The large filled symbols correspond to the detected sources,
 while the small triangles indicate the upper limits on the abundance of \HOCO.
In the panel with X[H$^{13}$CO$^+$], we do not show 05358-mm3, observed and detected
in \METH, \H\ and \HOCO\ but not in H$^{13}$CO$^+$.}
\label{fig_abb}
\end{center}
\end{figure}

\begin{figure}
\begin{center}
{\includegraphics[width=8.1cm,angle=0]{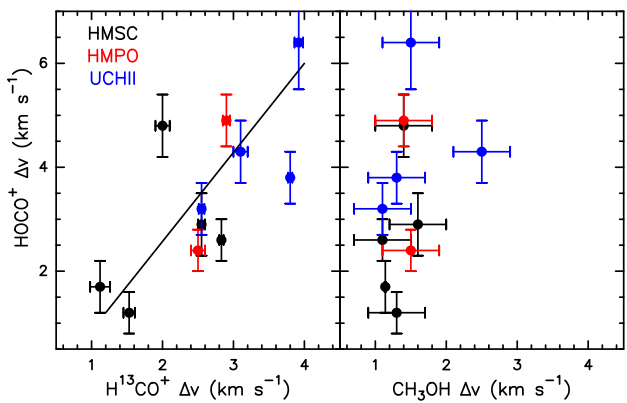}}
 \caption{Line widths at half maximum, $\Delta v$, of \HCOpI\ (left panel) and \METH\ (right panel) 
against those of \HOCO. The colours indicate the different evolutionary groups as in Fig.~\ref{fig_abb}.
The solid line in the left panel corresponds to a linear fit to the data.}
\label{fig_dv}
\end{center}
\end{figure}

As discussed in Sect.~\ref{intro}, two main pathways have been proposed for the formation of
\HOCO\ in dense gas: either the gas-phase reaction ${\rm HCO^+ + OH \leftrightarrow HOCO^+ + H }$, 
or the protonation of desorbed CO$_2$, mainly
upon reaction with H$_3^+$ (Vastel et al.~\citeyear{vastel}, Bizzocchi et al.~\citeyear{bizzocchi}). 
To investigate if and how our observational results can
put constraints on these alternative pathways, in Fig.~\ref{fig_abb} we show the fractional 
abundance of \HOCO\ (calculated as explained in Sect.~\ref{abundances}) against that of 
\HCOpI, the main precusor of \HOCO\ in the gas phase, and \METH, a tracer of surface chemistry.
For completeness, we also plot X[HOCO$^+$] against X[N$_2$H$^+$], because protonation
of CO$_2$ may occur also via reaction with \H.
 The \METH\ and \H\ column densities used to compute X[CH$_3$OH] and X[N$_2$H$^+$] 
have been taken from Fontani et al.~(\citeyear{fontani2015a}) and Fontani et al.~(\citeyear{fontani2015b}), 
respectively, and rescaled to the beam of the \HOCO\ observations. 
The \HCOpI\ column densities used to derive X[H$^{13}$CO$^+$] have been estimated 
from the integrated intensities of the \HCOpI\ 1--0 lines at 86754.288~MHz, serendipitously 
detected in the same dataset described in Sect.~\ref{obs} and in Colzi et al.~(\citeyear{colzi}). 
We have followed the same approach used for \HOCO, namely we assumed optically thin lines and 
LTE conditions (see Sect.~\ref{abundances}). We used the excitation temperatures listed in Table~1. 
The beam size is almost the same of that of \HOCO, hence all the fractional abundances in 
Table~1 are averaged over the same angular region.

Figure~\ref{fig_abb} indicates a clear non-correlation between the abundances of \METH\ and 
\HOCO, while \HCOpI\ and \HOCO\ seems positively correlated. By applying simple
statistical tests to the detected sources only, the correlation coefficient (Pearson's $\rho$) 
between X[\HOCO] and X[\HCOpI] is 0.7. Considering also the upper limits, the correlation remains positive
(Pearson's $\rho$ = 0.6). The correlation between X[\HOCO] and X[\H] is positive (Pearson's $\rho$ = 0.4 
without the upper limits) but much less convincing. 
If we assume that both CO$_2$ and \METH\ form on grain mantles, and what we find in
the gas is evaporated at similar times, the lack of correlation between X[\HOCO] and both
X[\METH] and X[\H] would indicate that the origin of \HOCO\ is likely not from CO$_2$ evaporated 
from ice mantles. This interpretation has two big caveats. First, the formation processes of CO$_2$
and \METH\ on the surfaces of dust grains can be different. In fact, CO$_2$ is thought to form 
in water ice mantles of cold carbonaceous grains via 
cosmic-ray bombardment (Mennella et al.~\citeyear{mennella}), or via the surface reaction CO + OH at 
dust temperatures of $\sim $30~K (Garrod \& Pauly~\citeyear{gep}), while \METH\ if formed from 
hydrogenation of CO at dust temperatures $\sim 10$~K (Vasyunin et al.~2017). 
Second, the main molecular ion responsible for the protonation of CO$_2$ is H$_3^+$ and not \H.
Nevertheless, the positive correlation between X[\HCOpI] and X[\HOCO] suggests a non negligible, 
or even dominant, contribution from \HCOp\ to the formation of the detected \HOCO. 
This interpretation of our results is in agreement with the study of 
Majumdar et al.~(\citeyear{majumdar}), who proposed that the dominant (up to 85$\%$) 
formation route of \HOCO\ in the extended and cold ($T\leq 30$~K) envelope of the hot corino 
IRAS 16293--2422 is indeed from the gas-phase reaction OH + \HCOp.
The fact that \HOCO\ and \HCOpI\ likely arise from similar gas can be understood
also from the comparison of their line widths at half maximum. Fig.~\ref{fig_dv} indicates that
the \HOCO\ line widths are correlated with those of \HCOpI, but not with those of \METH\ 
(which are always narrower). Hence, \METH\ is likely not associated with the same gas. 

Overall, our findings suggest a significant (perhaps dominant) role of HCO$^+$ as a gas-phase 
progenitor of \HOCO. However, caution needs to be taken in the interpretation of our results for several 
reasons. First, the \HOCO\ column densities, and hence the fractional abundances, have been derived
assuming an excitation temperature that could not be that of the molecule. To solve this problem, detection 
of more lines tracing different excitation conditions are absolutely required. Another caveat arises from 
the fact that our column densities are values averaged over large (28\asec) angular surfaces. Our targets 
are known to have complex structure, and temperature (and density) gradients. Therefore, higher angular 
resolution observations are needed to precisely determine the \HOCO\ emitting region, and, from this, 
understand its temperatures and densities, required to properly model the chemistry.

{\it Acknowledgments.}  We thank the IRAM-30m staff for the precious help during the observations.
We thank the anonymous referee for his/her constructive comments.
V.M.R. and M.P. acknowledge the financial support received from the European Union's Horizon 2020 research 
and innovation programme under the Marie Sklodowska-Curie grant agreement No 664931.
L.C. acknowledges support from the Italian Ministero dell'Istruzione, Universit\`a e Ricerca 
through the grant Progetti Premiali 2012 - iALMA (CUP C52I13000140001). P.C. acknowledges support 
from the European Research Council (ERC project PALs 320620).

\let\oldbibliography\thebibliography
\renewcommand{\thebibliography}[1]{\oldbibliography{#1}
\setlength{\itemsep}{-1pt}} 

{}

\end{document}